\documentclass[%
onecolumn,%
oneside,%
floats,%
aps,%
 prd,%
nobibnotes,%
nofootinbib,%
amsmath,%
amssymb,%
amsfonts,%
amscd,%
  superscriptaddress,%
eqsecnum%
]{revtex4}

\usepackage[utf8]{inputenc}
\usepackage{graphicx,array,dcolumn}
\usepackage{cases}
\usepackage[paperwidth=210mm,paperheight=297mm,centering,hmargin=1.8cm,vmargin=2.5cm]{geometry}
\usepackage{enumerate}

\usepackage{hyperref}

\usepackage[normalem]{ulem}
\usepackage{soul}

\def\fun#1#2{\lower3.6pt\vbox{\baselineskip0pt\lineskip.9pt
\ialign{$\mathsurround=0pt#1\hfil ##\hfil$\crcr#2\crcr\sim\crcr}}}

\newcommand{{\SD}}{\rm SD}

\newcommand{{\Mc}}{\mathcal{M}}

\newcommand{\vex}{\mbox{\boldmath${\rm x}$}}
 \newcommand{\ved}{\mbox{\boldmath${\rm d}$}}

\newcommand{\vesig}{\mbox{\boldmath${\rm \sigma}$}}

\newcommand{\veS}{\mbox{\boldmath${\rm S}$}}

\newcommand{\veR}{\mbox{\boldmath${\rm R}$}}

\newcommand{\veB}{\mbox{\boldmath${\rm B}$}}

\newcommand{\veE}{\mbox{\boldmath${\rm E}$}}

\newcommand{\vemu}{\mbox{\boldmath${\rm \mu}$}}

\newcommand{\ran}{\rangle}

\def\-g{\sqrt{-g}}

\newcommand{\be}{\begin{equation}}
\newcommand{\ee}{\end{equation}}

\newcommand{\ben}{\begin{equation*}}
\newcommand{\een}{\end{equation*}}

\newcommand{\bea}{\begin{eqnarray}}
\newcommand{\eea}{\end{eqnarray}}

\renewcommand\rho{\varrho}
\renewcommand\tilde{\widetilde}

\begin{document}

\title{\sc A quest for new physics inside the neutron\footnote{ Lecture at 44 th ITEP Winter
School}}

\author{\firstname{B.~O.}~\surname{Kerbikov}}

\email{borisk@itep.ru}

\affiliation{A.I.~Alikhanov Institute for Theoretical and Experimental Physics,
Moscow 117218, Russia}

\affiliation{Lebedev Physical Institute,
Moscow 119991, Russia}

\affiliation{Moscow Institute of Physics and Technology, Dolgoprudny 141700, Moscow Region, Russia}

\date{ }

\begin{abstract}
The lecture presents an overview of the quest for the new physics in low energy
neutron phenomena. In addition to the traditional topics the quantum damping of
$n\bar n$ oscillations is discussed.
 \end{abstract}

\maketitle

\section{Introduction \label{intro} }

The lecture has been delivered for young physicists and students
 with only basic knowledge of neutron physics.
  There are obvious caveats for experts associated with this.
  For the same reason the list of references is in no way intended to be complete.

Neutron was discovered by James Chadwick -- in 1932.  Its main characteristics
-- mass, lifetime, magnetic moment, may be found in Particle Data Group (PDG)
Reviews. It might seem  strange, but neutron lifetime is still a controversial
subject. This will be the topic of the next section. Probably the main
breakthrough in the new physics associated with neutron is the discovery of
parity nonconservation observed in 1956 by Chien-Shiung Wu group following the
theoretical physicists Tsung-Dao Lee and Chen-Ning Yang idea. The experiment
monitored   the $\beta$ decay of $^{60} Co$ $\left( ^{60} Co \to ^{60} Ni +
e^{-} + {\bar \nu}_{e}, \text{ or } d \to u + e^{-} + {\bar \nu}_{e} \text{ at
the quark level} \right)$. The $^{60} Co$ nuclei were aligned along the
direction of an external magnetic field. The decay probability may be written
as \be W = W_0 \left[ 1 + a_{Sp}({\vec S}_N{\vec p}_e) \right] \label{eq01} \ee
where ${\vec S}_N$ in the $^{60} Co$ spin, ${\vec p}_e$ is emitted electron
momentum. The last term in (\ref{eq01}) implies parity nonconservation since
spin does not change under reflection while momentum changes its sign. If
parity conservation were true in $\beta$-decay, electron would have no
preferred direction relative to ${\vec S}_N$ and the last term in (\ref{eq01})
would be equal to zero. Experimentally electrons were preferentially emitted in
the direction  opposite to that of the nuclear spin.

Not all topics covered in the  lecture may be attributed to ``new physics'',
``Beyond the Standard Model''. For example, neutron lifetime puzzle may be
either an experimental artifact, or the manifestation of mirror particles and
mirror magnetic field. The quantum damping phenomenon presented at the end of
the lecture is at the intersection of quantum mechanics and statistical
physics. This effect may play  a crucial role in future searches of
neutron-antineutron oscillations, and, more generally, in flavor, or
matter-antimatter oscillations.

The lecture is organized as follows.  In the next section we discuss the
neutron lifetime problem. Section 3 is devoted to the search for EDM of
neutron. Section 4 contains the discussion of the CP symmetry in $n-{\bar n}$
oscillations. In Section 5 the quantum damping phenomenon is presented.

\section{Neutron Lifetime: Beam VS Bottle}

There is a long-standing controversy of  experimental results for the neutron
lifetime $\tau$ obtained by the two complementary methods: the beam and the
storage (bottle) ones. The detailed description of the present situation and
the historical retrospective may be found in three recent presentations
\cite{01,02,03}.

Neutron lifetime $\tau$ is an important  quantity for the CKM Unitarity test
since $\tau^{-1}$ is proportional to $\left|V_{ud}\right|^2$. The theoretical
uncertainly of $\left|V_{ud}\right|^2$ is $4 \cdot 10^{-4}$ and this imposes
the requirement $\Delta\tau/\tau < 10^{-3}$. The primordial helium production
is very sensitive to the value of $\tau$ and this brings a constraint on the
baryon-to-photon ratio.

In the beam method protons from beta-decay of  cold neutrons $(v \sim 10^3
\text{ m/s})$ during its flight are counted. Absolute neutron flux must be
measured very accurately. The beam method may overestimate $\tau$ if a fraction
of beta-decay products is not detected. The best accuracy with the beam method
was achieved in \cite{04} \be \tau = 887.7 \pm 2.3\text{ s.} \label{eq02}\ee
The current  value based on two beam experiments is $\tau = 888.0 \pm 2.1\text{
s.}$ \cite{05}.

The bottle method may be described as: fill-store-count.  Use  is made of
ultracold neutrons (UCN, $v \sim 4 \text{ m/s}$) trapped in a closed volume.
The number of survived neutrons is measured as a function of time. The idea
that UCN undergo a complete reflection from the trap material is due to
Ya.B.~Zeldovich (1959). In reality, a small fraction of UCN may be lost due to
poorly controlled mechanisms. Therefore this method may underestimate the value
of $\tau$. Speaking about the bottled experiments, reference is most often is
made to \cite{06} \be \tau = 878.5 \pm 0.8\text{ s.} \label{eq03}\ee The most
recent storage value is \cite{07} \be \tau = 880.2 \pm 1.2\text{ s.}
\label{eq04}\ee The current average result for the bottled experiments is
$\tau= 879.6 \pm 0.8\text{ s}$ \cite{05}. Worth mentioning the preliminary
result of Serebrov group with big gravitrap: $\tau= 875.9 \pm 1.5\text{ s}$
\cite{01}. Recently the first UCN magnetic storage result was obtained
\cite{08} \be \tau = 878.3 \pm 1.9\text{ s.} \label{eq05}\ee The idea that low
energy neutrons may be confined by magnetic field goes back to V.V.~Vladimirsky
(1961).

From the above numbers it is clear that there is  a severe discrepancy between
beam and storage results. Loosely speaking, the standard Big Bang
nucleosynthesis seems to favor beam results. Particle physics experiments tend
to be in favor of the storage  results. Ambitious plans for future precision
measurements may be found in \cite{01,02,03}.

One may ask a question whether the above discrepancy  may be a manifestation of
some ``new physics'' playing a role only in one of the above methods. Here
comes the hypothesis \cite{09} that inside a trap neutrons may undergo
oscillations into mirror neutrons which freely leave the trap. More than that,
neutron-mirror-neutron oscillation frequency is sensitive to the direction of a
hypothetical mirror magnetic field. The idea of mirror world was introduced by
T.D.~Lee and C.N.~Yang in 1956 and developed by I.Yu.~Kobzarev, L.B.~Okun and
I.Ya.~Pomeranchuk in 1966. Necessary to note, however, that for the proposal
\cite{09} to work the oscillation time $\tau_{n n'}$ has to be as small as a
few seconds. Present experimental lower limit on $\tau_{nn'}$ is $\tau_{nn'} >
414 s$ \cite{10}. According to \cite{09} this result does not exclude the
oscillation time to mirror neutrons of the order of  a few seconds if the
presence of  mirror magnetic field is also assumed.

\section{The quest for EDM of the neutron}

During the last couple of years the new physics hopes at LHC have been fading
fast. On the other hand, we observe the revival of the interest to possible
magnifestations of BSM (Beyond the Standard Model) in low energy physics. There
is one effect the search for which lasts already for more than half a century.
This is  the  electric dipole moment (EDM) of the neutron. The possibility of
electric dipole moments for elementary particles was raised by E.M.Purcell and
N.F.Ramsey in 1950: ``The validity of  $P$ must rest on experimental
evidence''. These great authors proposed neutron--beam resonance experiment for
the detection of EDM.

The existence of neutron EDM would violate the CP--symmetry. To see this,
consider the non-relativistic neutron Hamiltonian in an electromagnetic field

\be \hat H =- \ved_n \veE - \vemu_n \veB,\label{6}\ee where  $\ved$ and $\vemu$
are the electric  dipole and magnetic moments correspondingly (we use the
natural system of units $\hbar = c =1, ~~ \alpha= e^2/4\pi$). The only vector
characteristic of the neutron is its spin $\veS$, and therefore both $\ved$ and
$\vemu$ should be either parallel, or antiparallel to $\veS$. Under space and
time inversion $\veS, \veE$ and $\veB$ transform as \be P: \veS \to \veS, ~~
\veE \to - \veE, ~~ \veB\to \veB,\label{7}\ee

\be t: \veS \to -\veS, ~~ \veE \to  \veE, ~~ \veB\to -\veB .\label{8}\ee

Therefore the term $\ved \veE \sim \veS \veE$ violates  both $P$- and $T$-
symmetries, while the magnetic coupling given by the second term defines the
neutron magnetic moment, $\mu_n =-1.91 \mu_N, ~~ \mu_N = e/2m_p$. The
observation of $ d_n$ would not only indicate the violation of $P$ and $T$ but
the $CP$  violation as well   since any locally Lorentz--covariant theory with
spin-statistics relation is $CPT$ invariant. Since its discovery in $K$-decays
in 1964, the $CP$ violation has been thoroughtfully studied both experimentally
(in particular in $B$-meson decays) and theoretically. The $CP$ violation is
 one of the three Sakharov conditions for Baryogenesis, in  other words, for
the explanation of matter-antimatter asymmetry in the Universe.

The neutron EDM is commonly measured in $e\cdot$ cm. This correspond to a naive
picture of two opposite charges separated by a distance $r$(cm). We refer to
recent reviews  \cite{11, 12} for the description of the past, present and
planned experiments aimed  at the detection  of the neutron EDM. The present
upper limit on the neutron EDM is 3.0$\cdot 10^{-26}  e\cdot$ cm  (90\% C.L.)
\cite{13,14}. The measurement rests upon two basic elements : a) the use of UCN
(the idea of F.L.Shapiro, 1969-1970), and b) the resonant frequency technique
(N.F.Ramsey, 1950). For the detailed description see \cite{11,12}. Several new
projects are under way.

In the SM the neutron EDM appears via the so-called penguin diagram \cite{15}
but its  expectation value is very small $\sim 10^{-32}  e\cdot$ cm. Another
possibility is to introduce into the QCD Lagrangian the $\theta$-term which is
odd under  time reversal and thus breaks CP \be L_\theta =
\frac{\theta}{16\pi^2} \varepsilon^{\mu\nu \alpha\beta} G^a_{\mu\nu}
G^a_{\alpha\beta} = \frac{\theta}{16 \pi^2} G\tilde G,\label{9}\ee where
$G^a_{\mu\nu}$ is the gluon field tensor, $\tilde G$ is its dual
 \be G^a_{\mu\nu} =\partial_\mu A^a_\nu - \partial_\nu A^a_\mu + ig f^{abc}
 A^b_\mu A^c_\nu.\label{10}\ee
 The $\theta$-term is similar to $(\veE \veB)$ product in electrodynamics.

 However,  from the experimental upper limit on $d_n$ the value of $\theta$
 should be tiny $\theta< 10^{-10}$. This situation  is called  the  ``Strong CP
 problem''. We do not discuss $d_n$ within the supersymmetric theories partly
 because no footprints of SUSY have been observed up to now.

 Recently \cite{16} a possibility that in strong magnetic field the induced
 neutron EDM is not generally constrained  to lie along its spin has been
 discussed.

\section{${\mathbf{n \bar n}}$ - oscillations and CP violation }

The observation of neutrons turning into antineutrons would constitute a
discovery of fundamental importance for particle physics and cosmology. It
would show that matter containing neutrons is unstable. The problem of $n\bar
n$-oscillations rises a lot of questions, both theoretical and experimental.
The recent review of its theoretical status and experimental prospects may be
found in \cite{17}.

The physical motivations for the search of $\Delta B  \neq 0$ process is
threefold:

\begin{description}
    \item[(i)]  Matter--antimatter asymmetry in the Universe,
    \item[(ii)] $B$-conservation is ``accidental'', no local $U(1)_B$ group
    unlike charge  conservation with $U(1)_{\rm em}$,
    \item[(iii)]  In SM $B$ is conserved only perturbatively  -- sphaleron
    breaks $B$ but conserves $(B-L)$.
\end{description}

In this section we discuss only one side of the general $n\bar n$ oscillations
problem, namely the possibility that oscillations imply the CP violation. This
question was raised in \cite{18} and  followed by a vivid discussion
\cite{19,20,22}. We start by quoting the conclusions from \cite{18} and
\cite{19}: ``neutron--antineutron oscillation implies breaking of CP along with
baryon number violation'' \cite{18},  and ``the neutron--antineutron
oscillation per se does not necessarily imply CP violation'' \cite{19}. Below
we follow in an oversimplified form the  arguments presented in \cite{19, 20}.
A reasonable assumption is that Lorentz invariance and CPT are valid in a
healthy field theory. The baryon number and parity symmetries are violated
since $\Delta B=2$ and parities of $n$ and $\bar n$ are opposite. The
lagrangian with $\Delta B =2$ term may be written in the following form

$$   \mathcal{L} = \bar n (x) i \gamma^\mu \partial_\mu n(x) - m \bar n (x)
n(x) -$$ \be -\frac12 \left[ \varepsilon \bar{ n^c} (x) n(x) + \varepsilon^*
\bar n (x) n^c (x) \right].\label{11}\ee

The charge conjugation is defined by \be \psi^c (x) = C\bar \psi^T(x), ~~ \bar
\psi^c (x) =  \psi^T(x) C, ~~ C= i\gamma^2\gamma^0.\label{12}\ee

The parity transformation is chosen as \be \Psi (t, \vex) \to \gamma_0 \Psi (t,
-\vex), ~~ \bar\psi (t, \vex) \to \bar \psi (t, -\vex)\gamma_0.\label{13}\ee

One may call this transformation $\gamma^0$ -- parity in contrast to
$i\gamma^0$ -- parity used in some  textbooks. In principle, one can add an
extra mass term $\left( - m' \bar n i \gamma_5 n\right)$ which we discard, see
\cite{18,19,20,21}.

The first term in (\ref{11}) is invariant under $P$, $C$ and $T$. The second
$\Delta B = 2 $ term is $P$ --  odd.

What about the $C$ -- parity of  this term? It is instructive to write
$\varepsilon$ in the form $ \varepsilon = | \varepsilon| e^{i\alpha}$ with real
$\alpha$. Then for $\alpha =0$ one  retrieves the $\Delta B =2$ Lagrangian of
\cite{18} which is obviously $C$ even and thus CP odd. However, for $\alpha=
\pi/2$ the second term in (\ref{11}) is $C$  odd and CP even. Physics seem to
depend on phase rotation! In other words, the CP  property of $ \Delta B = 2 $
term is ill--defined \cite{19}. The deeper insight into the problem requires
rather elaborate technique: Majorana fermions,  inclusion of external fields,
the spin dependence, Bogolyubov transformation, etc. \cite{20,22}. The general
conclusion remains the same: the appearance  of $n\bar n$ oscillations does not
in itself break CP.

\section{ Quantum damping of $\mathbf{ n\bar n}$  oscillations}

The lower limit in $n\bar n$ oscillation time was  set long ago  in the ILL --
Grenoble reactor experiment \cite{23}, $\tau_{n\bar n} > 0.86 \cdot 10^8s$
(about 3 years), or $\varepsilon$$ = \tau_{n\bar n}^{-1}  < 10^{-23}$ eV. The
internuclear experiments confirm this result up to a certain uncertainty  due
to the nuclear structure factors \cite{17}. The ``long'' $\beta$-decay lifetime
of the neutron $\lambda^{-1} = \tau_n = 0.88\cdot 10^3s$ (see Sec. II above) is
5 orders of magnitude  less than $\tau_{n\bar n}$ In the energy scale
$\varepsilon$ is more than 16 orders of magnitude less than the Lamb shift in
the hydrogen atom and about 10 orders of magnitude less than the hydrogen atom
bouncer energy \cite{24}. It is more correct therefore to use the term ``rare
decay'' than oscillation.

What can lead to an additional suppression of this rare process? First, this is
the splitting of $n$ and $\bar n$ states due to external magnetic  field,
$\omega = 2|\mu_n| B \simeq 6\cdot 10^{-12}$ eV for the Earth magnetic field
and $\omega   \simeq   10^{-15}$ eV for $B\sim  10nT$ in the experiments
\cite{10}. The second suppression factor estimated in \cite{10} as $\delta \sim
10^{-15} $ eV is due to the interaction with the residual gas. At this moment
we discard  this factor and will return to it shortly. With magnetic field
included the expression for $n \bar n$ oscillations reads \cite{17}. \be |
\psi_{\bar n} (t) |^2 = \frac{ 4 \varepsilon^2}{\omega^2 + 4 {\varepsilon}^2 }
e^{-\lambda t} \sin^2 \left( \frac12 \sqrt{ \omega^2 + 4 {\varepsilon}^2}~ t
\right).\label{14}\ee

Some refinements of this equation, like the wave packet formalism, may be found
in \cite{26, 27, 28}. The  free-space regime $| \psi_{\bar n} (t) |^2 \simeq
{\varepsilon}^2 t^2$ is `` hidden'' in (\ref{14}) in the short time limit $t\ll
2/\omega$ provided $\varepsilon$$ \ll\omega/{2}$. For $B\simeq  10^{-9} T$
foreseen in future experiments one has $2/\omega \simeq 0.2 s$, $ \omega/2
\simeq 3\cdot 10^{-14}$ eV $\gg\varepsilon$. We note in passing that the
$\varepsilon$$^2 t^2$ law does not allow to define the transition probability
per unit time.

 Now we consider in a schematic way the influence of environment on $n \bar n$
 oscillations. It is possible to develop a general density  matrix approach to
 $n\bar n$ oscillations when the role of the  ambient medium is played by the
 trap walls, nuclear  matter, or the residual gas inside the experimental
 setup. Here we consider only the residual gas case. Environment breaks the
 coherence \cite{29} of the propagation making the description in terms of the
 wave function impossible. The problem of oscillations in a gas was first
 solved using the density matrix in \cite{30} where muonium to antimuonium
 conversion was analyzed. For $n \bar n$ case similar approach has been
 developed in \cite{25}. The two-state system is described by the density
 matrix.

\be \hat\rho = \left( \begin{array}{cc}
 \varphi_1 \varphi_1^{*} & \varphi_1 \varphi_2^{*}\\
\varphi_1^{*} \varphi_2 & \varphi_2\varphi^{*}_2 \end{array} \right) = \left(
\begin{array} {ll}\rho_{11}& \rho_{12}\\
\rho_{21}& \rho_{22}
\end{array}\right).
\label{15}\ee

In vacuum and without decay $\hat \rho$ satisfies the von Neumann-Liouville
equation \be i\frac{d\hat \rho}{dt} = [\hat H, \hat \rho], \label{16}\ee

\be \hat H = \left( \begin{array} {cc} E+\Delta_1& \varepsilon\\ \varepsilon&
E+ \Delta_2\end{array}\right), \label{17}\ee where $\Delta_1-\Delta_2 =
\omega$. Equation (\ref{16}) may be represented in a vector form of the Bloch
equation \cite{31}. The real Bloch vector $\veR$ is introduced by the expansion
of the density matrix over the Pauli matrices

\be \hat \rho = \frac12 (1+ \veR \vesig).\label{18}\ee Then \be \veR = \left(
\begin{array}{l} \rho_{12}+ \rho_{21}\\-i(\rho_{21}+ \rho_{12})\\\rho_{11}+
\rho_{22}\end{array} \right).\label{19}\ee

\newcommand{\veV}{\mbox{\boldmath${\rm V}$}}
The von Neumann-Liourille equation (\ref{16}) may be written in the following
form \be \dot{\veR} = \veV \times \veR, \label{20}\ee

\be \veV = \left( \begin{array}{l} 2
\varepsilon\\0\\\omega\end{array}\right).\label{21}\ee

Equation (\ref{20}) describes the precession of $\veR$ around the ``magnetic
field'' $\veV$. Equations (\ref{16}) and (\ref{20}) does not include
decoherence since they correspond to an isolated system. As before, in the
short time limit they yield $\rho_{22} \simeq \varepsilon^2 t^2$. Interaction
with the environment destroys the off-diagonal elements of $\hat \rho$ and this
is the essence of decoherence. As a result the interference between the two
basic states $|n \ran$ and $|\bar n \ran$  may become impossible.

Below we present an oversimplified scenario of oscillations damping. Let
$\tau_i$ be the time interval between the $(i-1)-$th  and  $i-$th collisions
with the gas molecules. We also introduce the average time between collisions
$\tau= t/n$, $n$  is  the number of  collisions, $t$ is the observation time.
In this  simple picture we do not care that in a real experiment $n$ may be of
the order of one, we only assume that $\varepsilon \tau \ll 1, $ i.e., $ \tau
\ll 10^8 s$. To  make things even simpler we ignore the external magnetic field
and put $\omega =0$. According to (\ref{20}) and (\ref{21}) the evolution of
the system before the first collision proceeds according to \be \dot{R}_z =  2
\varepsilon R_y, ~~\dot{R}_y =  -2 \varepsilon R_z.\label{22}\ee Then just
before the first collision (\ref{22}) yields \be R_z(\tau_1) = \cos 2
\varepsilon \tau_1, ~~R_y(\tau_1) = -\sin 2 \varepsilon \tau_1.\label{23}\ee

At the collision the $\bar n$ component gets annihilated while the $n$
component is assumed to  stay intact. Then just after the collision one has \be
R_z(\tau_1) = \cos^2 \varepsilon \tau, ~~R_y(\tau_1) = 0.\label{24}\ee

Note  that $\tau_1$ in (\ref{23}) and (\ref{24}) differ by the collision time
which is discarded  here. Under the assumption $\varepsilon \tau \ll 1$ and
averaging over the time  intervals between collisions one obtains \cite{25, 32}
\be R_2 =\prod^n_{k=1} \int^\infty_0 \frac{d\tau_i}{\tau} \exp \left( - \frac{
\tau_i}{\tau}\right) cos^2 (\varepsilon \tau_i) \simeq \exp (- 2 \varepsilon^2
\tau t). \label{25}\ee

At this point we note  that one can arrive at the result (\ref{25}) if an
additional damping parameter $\rho$ is introduced into (\ref{22}), namely \be
\dot{R}_z= 2 \varepsilon R_y, ~~\dot{R}_y =  -2 \varepsilon R_z- \rho
R_y.\label{26}\ee The factor $\rho$ should not be confused with the
$\beta$-decay parameter $\lambda = 1/\tau_\beta$ which enters into the
equations for all three components of $\veR$ on equal footing ($\beta$-decay is
temporary discarded). From (\ref{26}) one obtains the following equation for
$R_z$ \be\ddot{ {R}}_z + \rho \dot{R}_z +4 \varepsilon^2R_z =0.\label{27}\ee

In the limit $\rho\gg \varepsilon$ the solution of (\ref{27}) has the form \be
R_z \simeq \exp \left( -\frac{4\varepsilon^2}{\rho} t \right).\label{28}\ee

The  condition to match (\ref{25}) is \be \rho = \frac{2}{\tau} \sim \nu v
\sigma_a, \label{29}\ee where $\nu$ is the number density of the residual gas
molecules, $v$ is the mean velocity between $n$ and the gas molecules,
$\sigma_a$ is the annihilation  cross section. For $\rho\gg \varepsilon$ the
solution (\ref{28}) means that at any time \be \left| \frac{ \psi_{\bar n}
(t)}{\psi_n(t)}\right|^2 \simeq \frac{4\varepsilon^2}{\rho^2} \ll
1.\label{30}\ee In terms of the Bloch vector the overdamping regime (\ref{28})
means that due to annihilation $R_z$ does not have enough time to turn from
$R_z=1, \rho_{11}=1, \rho_{22}=0$ to $R_z=-1 , \rho_{11}=0, \rho_{22}=1$.

 The overdamping regime previously discussed in \cite{25,33} and
within the general theory of decoherence in \cite{29}. Whether this regime
might be of importance in already performed and planned experiments is a topic
of  a separate  investigation.

The decoherence caused by the interaction with the environment is a general
phenomenon. It takes place in neutrino oscillations \cite{34}, positronium
\cite{35} and neutron{\cite{36} oscillations to mirror, or brane worlds, heavy
quarks oscillations in the  color gluon field environment  \cite{37}, $B$-and
$K$-mesons oscillations \cite{38, 39}.

 \vspace{0.5cm}
 \noindent
 {\bf Acknowledgement}\\
 This work was supported by the grant
   from the Russian Science Foundation project number 16-12-10414. The author express his deep
  gratitude to Y.Kamyshkov for numerous illuminating discussion
  and suggestions  and to V.Novikov for useful remarks.








\end{document}